\documentstyle[epsf,onecolumn]{mn}

\def\centerline#1{\hbox to\hsize{\hss#1\hss}}
\makeatletter
\def\eqalign#1{\null\,\vcenter{\openup\jot\m@th
  \ialign{\strut\hfil$\displaystyle{##}$&$\displaystyle{{}##}$\hfil
      \crcr#1\crcr}}\,}
\def\eqalignleft#1{\null\,\vcenter{\openup\jot\m@th
  \ialign{\strut$\displaystyle{##}$\hfil&$\displaystyle{{}##}$\hfil
      \crcr#1\crcr}}\,}
\makeatother
\begin{document}

\title{Formation and coalescence of relativistic binary stars:
effect of kick velocity}

\author[V.~M.~Lipunov, K.~A.~Postnov and M.~E.~Prokhorov]
       {V.~M.~Lipunov, K.~A.~Postnov and M.~E.~Prokhorov\\
Sternberg Astronomical Institute, 119899 Moscow, Russia}
\date{Accepted 1996 ...
      Received 1996 ...}

\pagerange{\pageref{firstpage}--\pageref{lastpage}}
\pubyear{1996}

\maketitle

\baselineskip=1.6\baselineskip

\label{firstpage}

\begin{abstract}

For the first time, using Monte-Carlo calculations of the modern
scenario for binary stellar evolution with account for spin evolution
of magnetized compact stars (the ``Scenario Machine''), we compute the
amount of galactic binary pulsars with different companion types
(OB-star, white dwarf (WD), neutron star (NS), black hole (BH), or
planet) assuming various phenomenological distributions for kick
velocities of newborn NS.  We demonstrate a strong dependence of the
binary pulsar population fractions relative to single pulsars on the
mean kick velocity and found an optimal kick velocity of 150-200 km/s.

We also investigate how the merging rates of relativistic binary stars
(NS+NS, NS+BH, BH+BH) depend on the kick velocity.  We show that the
BH+BH merging may occur, depending on parameters of BH formation, at a
rate of one per 200,000 -- 500,000 years in a Milky Way type galaxy.
The NS+NS merging rate $R_{ns}$ is found to be 1 per $\sim 3,000$ years
for zero recoil, and decrease to one per $10,000$ years even for the
highest kick velocities of 400 km/s.

That the merging rates derived from evolutionary calculations are by
two order of magnitude higher than those based on binary pulsar
statistics only, is suggested to be due to the fact that the observable
binary pulsars in pairs with NS form only a fraction of the total
number of binary NS systems.

The merging rates obtained imply the expected detection rate of binary
BH by a LIGO-type gravitational wave detector comparable with and even
higher than the binary NS merging rate for a wide range of parameters.
Detecting the final frequency of a merging event at about 100 Hz and
the shaping of the waveforms would bring a firm evidence of BH
existence in nature.


\end{abstract}

\section{Introduction}

Among about 700 known radiopulsars (rapidly rotating, magnetized NS),
several tens are observed in binary systems (Taylor et al. 1993). The
secondary companion to a pulsar in a binary system can be one of most
classes of known celectial bodies: another NS (Hulse \& Taylor 1975), a
white dwarf (Wright \& Loh 1986), a giant blue star (Johnston et al.
1992), or a planet (Wolszcan \& Frail 1992).

The formation of a NS during the supernova explosion is usually
accompanied by a catastrophic mass loss, which in most cases leads to
the binary system disruption; a young NS can thus "remember" the
orbital velocity of the progenitor star before the collapse of the
order of a few 100 km/s (``Blaauw mechanism'', Blaauw 1961; Gott et al
1970).  However, the range of stellar parameters (masses, radii,
orbital separations etc.) is so wide that some part of the systems must
survive as binaries during the cataclysmic processes of stellar
collapse (see Bhattacharia \& van den Heuvel 1991). In the
framework of the standard scenario of binary system evolution (see van
den Heuvel 1994 for a review) diverse species of pulsars are produced
and the mean velocities of radiopulsars of about 100-200 km/s that were
measured shortly after their discovery (Manchester \& Taylor 1977) can
be naturally explained by high mass loss during the supernova explosion.

Resent revision of the pulsar distance scale has led to two-fold increase
in the mean pulsar transverse velocities derived from pulsar proper motion
(Lyne \& Lorimer 1994):  they now become as
high as 400 km/s.  Independently, even higher pulsar birth velocities of
$\sim 500-900$ km/s has been found from studying young pulsar
positions inside the associated supernova remnants (Frail et al.
1994).  To explain this,  the idea of an asymmetrical supernova
collapse proposed by Shklovskii (1970) has been reanimated.

Since Lyne \& Lorimer's distribution for pulsar transverse
velocities is based on the distance scale model, it may be
controversial and it has been suggested that
no natal kick velocity is needed to
fit the pulsar velocity data (see, e.g. Iben \& Tutukov 1996).  In the
present paper we show, using quite independent arguments, that the
surprising diversity of binary pulsar companions and the presently
observed number of such binary pulsars, strongly contradicts to the
hypothesis of the high birth pulsar velocities, generally because such
velocities would result in a much more effective disruption of binary
systems. However, in contrast to Iben \& Tutukov (1996), we show
that the observational data does require a kick velocity distribution with
the mean value of about 150 km/s and a long power-law high-velocity tail.

We further calculate the effect of the kick velocity on the merging
rate of binary relativistic stars (NS+NS, NS+BH, BH+BH), which are
among the primary targets for gravitational wave observatories
currently under construction (LIGO, VIRGO, GEO-600) (Abramovici et al.
1992; Schutz 1996).  Such systems are observed in the Galaxy if one of
the components shines as a radiopulsar.  Several binary pulsars with
massive compact secondary companions are known to date:  PSR 1913+16,
PSR 1534+12, PSR 0655+64, PSR 2303+46 (see Table 18 in van den Heuvel 1994)
and recently discovered PSR J1518+4904 (Nice, Sayer \& Taylor 1995), of
which two first have orbital periods $\le 10$ hr and will merge due to
gravitational radiation losses within less than the age of the Universe
$\sim 15\times 10^9$ years.

It is very important to know the accurate rate of such events, as the
planned LIGO sensitivity will allow detection of NS+NS mergings out to
$\sim 200$ Mpc (Abramovici et al. 1992). The problem seems even more
actual in view of modern cosmological models of gamma-ray
bursts involving binary NS or NS+BH coalescences (Blinnikov et al.
1984, Lipunov et al. 1995).  Attempts to estimate this rate ($R_{ns}$)
in our Galaxy have been constantly made over last 20 years using
theoretical considerations (Clark et al.  1979, Lipunov et al. 1987,
Narayan et al.  1991, Tutukov \& Yungelson 1993, Lipunov et al. 1995,
Dalton \& Sarazin 1995) and binary pulsar observations
(Phinney 1991, Curran \& Lorimer 1995).  The estimations based only on
the existing binary pulsar statistics can place an
ultraconservative lower limit of 1 per $\sim 10^6$ year, whereas those
based on theoretical evolutionary considerations always produce two
order of magnitude higher rates of 1 per $10^4$ year (see discussion
in van den Heuvel 1994).  Obviously, the difference comes from the fact
that, by deriving the merging rate from pulsar statistics, one should
rigorously take into account different selection effects (such as
pulsar beaming, completeness of pulsar surveys, possibility of both
components to be non-pulsars, etc.). Clearly, an accurate account for
the spin evolution of magnetized NS is needed here.

A criticism of the theoretical evolutionary estimates
is usually made with the reference to a large number of
poorly determined parameters of the evolutionary
scenario for massive binary systems, such as common envelope stage
efficiency, initial mass ratio distribution, distribution of the recoil
velocity imparted to NS at birth, etc. (Lipunov et al. 1996).  However,
by comparing the results of the Scenario Machine calculations with
other observations (Lipunov, Postnov, Prokhorov (hereafter LPP)
1996a,b) we may fix some free parameters (such as the form of the
initial mass ratio distrbution and the common envelope efficiency), and
then examine the dependence of the double NS merging rate on one
free parameter, say the mean kick velocity value. In fact, the calculations
turn out to be most sensitive to just
the kick velocity (LPP 1996b) as the binary
system gets more chances to be disrupted during supernova explosion,
especially when the recoil velocity becomes higher than the orbital
velocity of stars in the system.

In this paper we also compute the ratio of binary NS with active pulsars to
the total number of binary NS, and find the dependence of the binary NS
merging rate on the recoil velocity acquired by NS at birth assuming different
distributions for this velocity. A recent study of such a dependence was
done by Portegies Zwart \& Spreew (1996); they, however, examined only a
Maxwellian kick velocity distribution, whereas the observations of Lyne \&
Lorimer (1994) imply a different law for kick velocity (see below).

A special attention is given to calculating
galactic NS+BH and BH+BH mergings ($R_{bh}$).
For zero kick velocity this rate is found
to be at least 1-2 orders of magnitude smaller than that of NS+NS
(eg. Tutukov \& Yungelson, 1993).
However, the formation of a BH in a binary system may be accompanied by
a smaller recoil and, hence, the kick velocity would have
a smaller effect on the NS+BH
merging rate. Thus we may expect a smaller difference between the two rates
if the mean kick velocity is as high as indicated
by pulsar proper motions (of order 400 km~s$^{-1}$ or more). This in turn
would imply that the BH+BH rate might be equally important
from the point of view of gravitational wave registration by the LIGO type
detectors, since, having at least a few time higher masses, BH binaries
may be observed from about ten times farther distances by a detector with
given sensitivity (characteristic dimensionless strain metric
amplitude from a merging binary system, $h_c$, scales as $M^{5/6}/r$,
where $M$ is a characteristic mass of binary companions and $r$ is a
distance to the source (Abramovici al. 1992)).  Therefore the number
of events registered by the detector scales as
\[
\label{rates}
\frac{N_{bh}}{N_{ns}}\approx\left(\frac{R_{bh}}{R_{ns}}\right)
                \left(\frac{M_{bh}}{M_{ns}}\right)^{15/6}
\]
and may well be of order unity for typical $M_{bh}\sim 10 M_\odot$.
We address this question in more detail in a separate paper
(Lipunov et al. 1997).

\section {The Model}

Monte-Carlo simulations has shown its
ability to study successfully the
evolution of a large ensemble of binaries  and to estimate the number
of binaries at different evolutionary stages.  This method has become
popular over last ten years ( Kornilov \& Lipunov 1984; Dewey \& Cordes
1987; Bailes 1989; for another applications of Monte-Carlo simulations
see de Kool 1992; Tutukov \& Yungelson 1993; Pols \& Marinus 1994).

For modeling binary evolution, we use the ``Scenario Machine'', a
computer code based on modern binary evolution scenarios (for
a review, see van den Heuvel (1994)), which also takes into account the
influence of magnetic field of compact objects on their observational
appearance. A detailed description of the computational techniques and
input assumptions is summarized elsewhere (LPP 1996a), and here we list
only the basic parameters and initial distributions.

\subsection{Initial binary parameters}

The initial  parameters determining binary evolution are: the
mass of the primary ZAMS component, $M_1$; the binary mass ratio,
$q=M_2/M_1<1$; the orbital separation, $a$. We assume zero initial
eccentricity.

The distribution of initial binaries
over orbital separations is known from observations
(Abt 1983):
\begin{equation}
\eqalign{&f(\log a) ={\rm const}\,,\cr
&\max~\{10~ {\rm R}_\odot,~\hbox{Roche Lobe}~
(M_1)\} < \log a < 10^4~{\rm R}_\odot.
\cr}
\end{equation}

The initial mass ratio distribution in binaries, being very crucial for
overall evolution of a particular binary system (Trimble 1983),
has not yet been reliably derived
from observations due to a number of selection
effects.  A `zero assumption' usually made is that the mass ratio
distribution has a flat shape, i.e. the high mass ratio binaries
are formed as frequently as those with equal masses (e.g. van den Heuvel 1994).
Ignoring the real distribution, we parametrized it by a power
law, assuming the primary mass distribution to obey the Salpeter mass
function (Salpeter 1955):
\begin{equation}
\eqalignleft{
\frac{dN}{dM_1}=0.9\hbox{yr}^{-1} M_1^{-2.35}\,,\quad &0.1~{\rm M}_\odot < M_1< 120~{\rm M}_\odot\,;
\cr f(q)   \propto  q^{\alpha_q}\,,\quad&q\equiv M_2/M_1<1 \,;   \cr }
\end{equation}

This distribution produces approximately 1 massive star
($M_1>10 $ M$_\odot$) per 60 year in a binary system
(assuming $50\%$ stars in the Galaxy in binaries),
which coincides with the binary birthrates
derived from observations (Popova et al., 1982)

A comparison of the observed X-ray source statistics with the
predictions of the current evolutionary scenarios indicates (LPP 1996a)
that the initial mass ratio should be strongly centered around unity,
($\alpha_q\sim 2$). Of course, this is not a unique way of
approximating the initial binary mass ratio (see e.g. Tout (1991)).
However, from the point of view of binary NS merging rate,
this parameter affects the results much less
than the kick velocity. In the present paper, we use both
$\alpha_q=2$ and $\alpha_q=0$.

\subsection{Initial parameters of compact stars}

We are interested in binary NS or NS+BH systems,
so it is enough to trace
evolution of binaries with primary masses $M_1>10 M_{\odot}$ which are
capable of producing NS and BH in the end of evolution.
The secondary component can have a mass
from the whole range of stellar masses $0.1 M_\odot<M_2<120 M_\odot$.

We consider a NS with a mass of $1.4~ {\rm M}_{\odot}$ to
result from the collapse of a star with the core mass prior to the
collapse $M_*\sim (2.5-35)~{\rm M}_{\odot}$.  This corresponds to an
initial mass range $\sim (10 - 60)~{\rm M}_{\odot}$, considering
that a massive star can loose more than $\sim (10-20)\%$ of its
initial mass during the evolution with a strong stellar wind (de Jager
1980).

The magnetic field of a rotating NS largely defines the evolutionary
stage the star would have in a binary system (Schwartzman 1970;
Davidson \& Ostriker 1973; Illarionov \& Sunyaev 1975).  We use a
general classification scheme for magnetized objects elaborated by
Lipunov (1992).

Briefly, the evolutionary stage of a rotating magnetized NS  in a
binary system depends on the star's spin period $P$ (or spin frequency
$\omega=2\pi/P$), its magnetic field strength $B$ (or, equivalently,
magnetic dipole moment $\mu=BR^3/2$, where $R$ is the NS radius), and
the physical parameters of the surrounding  plasma (such as density
$\rho$ and sound velocity $v_s$) supplied by the secondary star.  The
latter, in turn, could be a normal optical main sequence (MS) star, or
red giant, or another compact star). In terms of the Lipunov's
formalism, the NS evolutionary stage is determined by one or another
inequality between the following characteristic radii: the light
cylinder radius of the NS, $R_l=c/\omega$ ($c$ is the speed of light);
the corotation radius, $R_c=(GM/\omega^2)^{1/3}$; the gravitational
capture radius, $R_G=2GM/v^2$ (where $G$ is the Newtonian gravitational
constant and $v$ is the NS velocity relative to the surrounding
plasma); and the stopping radius $R_{stop}$. The latter is a
characteristic distance at which the ram pressure of the accreting
matter matches either the NS magnetosphere pressure (this radius is
called Alfven radius, $R_A$) or the pressure of relativistic particles
ejected by the rotating magnetized NS (this radius is called
Schwartzman radius, $R_{Sh}$).  For instance, if $R_l>R_G$  then the NS
is in the ejector stage (E-stage) and can be observed as a radiopulsar;
if $R_c<R_A<R_G$, then so-called propeller regime is established
(Illarionov \& Sunyaev 1975) and the matter is expelled by the rotating
magnetosphere; if $R_A<R_c<R_G$, we deal with an accreting NS
(A-stage), etc. These inequalities can easily be translated into
relationships between the spin period $P$ and some critical period that
depends on $\mu$, the orbital parameters, and accretion rate $\dot M$
(the latter relates $v$, $v_s$, $\rho$, and the binary's major semiaxis
$a$ via the continuity equation). Thus, the evolution of a NS in a
binary system is essentially reduced to the NS spin evolution
$\omega(t)$, which, in turn, is determined by the evolution of the
secondary component and orbital separation $a(t)$. Typically, a single
NS embedded into the interstellar medium evolves as $E\to P\to A$
(for details, see Lipunov \& Popov 1995). For a NS in a binary, the
evolution complicates as the secondary star evolves: for example, $E\to
P\to A\to E$ (recycling), etc.

When the secondary component in a binary fills its Roche lobe, the
rate of accretion onto the compact star can reach the value
corresponding to the
Eddington luminosity $L_{Edd}\simeq 10^{38}~(M/{\rm M}_\odot)$
\hbox{erg/s}\, at the $R_{stop}$; then a supercritical regime sets in
(not only superaccretors but superpropellers and superejectors can
exist as well; see Lipunov 1992).

If a BH is formed in due course of the evolution, it can only appear as
an accreting or superaccreting X-ray source; other very interesting
stages such as BH + radiopulsar which may constitute a notable fraction
of all binary pulsars after a starburst are considered in Lipunov et
al. (1994, 1995a).

The initial distribution of magnetic fields of NS is another
important parameter of the model.  This cannot be taken from studying
pulsar magnetic field (clearly, pulsars with highest and lowest
fields are difficult to observe).  In the present calculations we
assume a flat distribution for dipole magnetic moments of newborn NSs %
\begin{eqnarray} f(\log\mu)=~{\rm const}\,,~~ 10^{28} \le \mu \le
      10^{32}~ \hbox{G cm}^3\,,
\end{eqnarray}
and the initial rotational period of the NS is assumed to be $1$~ms.

The computations were made under different assumptions about the
NS magnetic field decay, taken in an exponential form,
$\mu(t)\propto \exp(-t/tau)$, where $\tau$ is the characteristic decay
time  of $10^7-10^8$ year.  The field is assumed to
stop decaying below a minimum value of $10^9$ G (van den Heuvel et al.
1986). No accretion-induced magnetic field decay
is assumed.

A radiopulsar was assumed to be turned ``on'' until its period $P$ has
reached a ``death-line'' value defined from the relation
$\mu_{30}/P_{death}^2=0.4$, where $\mu_{30}$ is the dipole magnetic
moment in units of $10^{30}$ G~cm$^3$, and $P$ is taken in seconds.

The mass limit for NS (the Oppenheimer-Volkoff limit) is $M_{OV}=2.5~
{\rm M}_\odot$, which corresponds to a hard equation of state of the NS
matter.  The most massive stars are assumed to collapse into a BH once
their mass before the collapse is $M>M_{cr}=35~ {\rm M}_\odot$ (which
would correspond to an initial mass of the ZAMS star $\sim 60~ {\rm
M}_\odot$ since a substantial mass loss due to a strong stellar wind
occurs for the most massive stars).  The BH mass is calculated as
$M_{bh}=k_{bh}M_{cr}$, where the parameter $k_{bh}$ is taken to be 0.3,
as follows from the studies of binary NS+BH (Lipunov et al.  1994).

\subsection{Other parameters of the evolutionary scenario}

The fate of a binary star during evolution mainly depends on the
initial masses of the components and their orbital separation.  The
mass loss and kick velocity are the processes leading to the binary
system disruption; however, there are a number of processes connected
with the orbital momentum losses tending to bound the binary (e.g.,
gravitational radiation, magnetic stellar wind).

\subsubsection{Common envelope stage}

We consider stars with a constant (solar) chemical composition.  The
process of mass transfer between the binary components is treated
according to the prescription given in van den Heuvel (1994) (see LPP
(1996a) for more detail). The non-conservativeness of the mass transfer
is treated via ``isotropic re-emission'' mode (Bhattacharya \& van den
Heuvel 1991).  If the rate of accretion from one star to another is
sufficiently high (e.g. the mass transfer occurs on a timescale 10
times shorter than the thermal Kelvin-Helmholz time for the normal
companion), or the compact object is engulfed by a giant companion, the
common envelope (CE) stage of the binary evolution  can set in (see
Paczy\'nski 1976; van den Heuvel 1983).

During the CE stage, an effective spiral-in of the binary components
occurs.  This complicated process is not fully understood as yet, so we
use the conventional energy consideration to find the binary system
characteristics after the CE stage by introducing a parameter
$\alpha_{CE}$ that measures what fraction of the system's orbital
energy goes, between the beginning and the end of the spiralling-in
process, into the binding energy (gravitational minus thermal) of the
ejected common envelope. Thus,
\begin{equation}
\alpha_{CE}\left({GM_aM_c\over 2a_f}-{GM_aM_d\over 2a_i}\right) = {GM_d
    \left(M_d-M_c\right) \over R_d}\,,
\end{equation}
where $M_c$ is the mass of the core of the mass loosing star of initial
mass $M_d$ and radius $R_d$ (which is simply a function of the initial
separation $a_i$ and the initial mass ratio $M_a/M_d$), and no
substantial mass growth for the accretor is assumed (see, however,
Chevalier 1993).  The less $\alpha_{CE}$, the closer becomes
binary after the CE stage. This parameter is poorly known
and we varied it from 0.5 to 10  during calculations.

\subsubsection{High and low mass-loss scenario form massive star
evolution}

A very important parameter of the evolutionary scenario is
the stellar wind mass loss effective for massive stars.
No consensus on how stelar wind mass loss occurs in
massive stars exist.
So in the spirit of our scenario approach
we use two "extreme", in a sense, cases. The "low mass-loss"
scenario  treats
the stellar wind from a massive star of luminosity $L$
according to de Jager's prescription
\[
\dot M \propto \frac{L}{cv_\infty}
\]
where $c$ and $v_\infty$ are the speed of light and of the stellar
wind at infinity, respectively. This leads to at most 30 per cent
mass loss for most massive stars.

The "high stellar wind mass-loss" scenario uses calculations of single
star evolution by Schaller et al. (1992).  According to these
calculations, a massive star lose most of its mass by stellar wind down
to 8-10 $M_\odot$ before the collapse, practically independently on its
initial mass.  In this case we assume the same mechanism for BH
formation as for the "low mass-loss" scenario, but only one parameter
$k_{bh}$ remains ($M_{cr}$ is taken from evolutionary tracks).  Masses
of BH formed within the framework of the high mass-loss scenario are
thus always less than or about of 8 M$_\odot$.

So far we are unable to choose between the two scenarios; however,
recently reported observations of a very massive WR star of 72 M$_\odot$
(Rauw et al. 1996) cast some doubts on very high mass-loss scenario or
may imply that different mechanisms drive stellar wind mass loss.
However, we shall use the "high stellar wind mass-loss" scenario
when studying binary BH formation rates.

\section{Phenomenological kick velocity}

Ozernoy (1965) and Shklovskii (1970) were among the first who noted
that the collapse of a normal star into a compact relativistic object
can be anisotropical and thus the stellar remnant can acquire a space
velocity much higher than that of the progenitor star, which is
typically of the order of a few tens km/s.  Due to an enormous energy
liberated during the collapse, which is comparable with the rest-mass
energy of the whole star, $\approx Mc^2$, a small anisotropy
$\alpha\simeq 10^{-6}$ would be sufficient for the remnant to leave the
Galaxy at all having the velocity $w=\sqrt{2\alpha}\,c$, where $c$ is
the speed of light. A number of the anisotropy mechanisms has been
proposed: asymmetric neutrino emission in a strong magnetic field
during the collapse (Chugai 1984, Bisnovatyi-Kogan 1993); double
neutron star formation during the core collapse (Imshennik 1992 );
tidally induced asymmetric ignition of the white dwarf during the
accretion induced collapse (Lipunov et al. 1987), etc.  Recent
modelling of the core collapse by Burrows, Hayes and Fryxell (1995)
makes the neutrino induced asymetry very promisive (see Burrows \&
Hayes 1995).  Thus, similar to the cosmological constant term, the
anisotropy was released away (like a jinnee from the bottle) as a
possible but not necessary thing.

\begin{table*}
 \centering
 \begin{minipage}{140mm}
\caption{Observational zoo of the field galactic binary pulsars}
\begin{tabular}{@{}lr@{\quad\quad}r@{$\,$\quad}p{5cm}r@{}}
Type
&\multicolumn{1}{r}{Number}
&\multicolumn{1}{r}{Fraction}
&\multicolumn{1}{c}{Assumed origin}
&Reference\\[10pt]
PSR+NS&4&0.7\%&
From massive main se\-qu\-en\-ce stars
($M_{(1,2)}>10M_\odot$) 
&Lorimer et al (1995)\\
PSR+WD&15&2.5\%&
From massive binary with
large initial mass ratio or
accretion induced WD-collapse in LMXB 
& Lorimer et al (1995)\\
PSR+PL&2&0.3\%&
From close NS+WD after
WD mass loss to $0.001 M_\odot$
by Roche lobe overflow 
&
Wolsczan$\,\,$\&$\,\,$Frail (1992); Shabanova T.V. (1995)
\\
PSR+OB&1&0.2\%&
After first SN explosion in
massive binary with radio
transparent stellar wind 
& Johnston et al (1992)\\
Single PSR&
$\approx$600&100\%&
From disrupted binaries
&Taylor et al (1993)
\end{tabular}
\end{minipage}
\end{table*}

It seems natural to assume that the kick velocity is
arbitrarily directed in space. The value of the kick velocity
$w$ can be quite different and may depend on some
parameters, such as the magnetic field strength, angular velocity,
and so on; first we consider two extreme
assumptions of a strongly determined distribution,
$f_\delta(w)\propto \delta(w)$ and of a maxwelian-like one,
\begin{equation}
f_m(w)\propto w^2 \exp(-w^2/w_0^2)\,,
\label{Mxw}
\end{equation}
which is natural to
expect if several independent approximately equally powerful
anisotropy mechanisms randomly operates.
Here $w_0$ is a parameter which is connected with the mean
kick velocity $w_m$ by a relation $w_m=\frac{2}{\sqrt\pi}w_0$.

Since the recent results of Lyne \& Lorimer
(1994) suggest different distribution, we tried to model the
cumulative distribution of the transverse pulsar velocities
given in their paper. We found that a \it 3D-distribution \rm
for the kick velocity imparted to a newborn pulsar of the form
\begin{equation}
f_{LL}(x)\propto \frac{x^{0.19}}{(1+x^{6.72})^{1/2}}
\label{LLkick}
\end{equation}
where $x=w/w_0$, $w_0$ is a parameter, fits well the  Lyne \& Lorimer's
2D-distribution at $w_0=400$~ km/s.  This distribution has a power-law
asymptotic behaviour at low velocities ($x\ll 1$), but goes flatter
($\propto x^0.19$) than the Maxwellian one ($\propto x^2$). As the kick
velocities of order or less than the orbital velocities of stars are
the most important for the binary system fate after the supernova
explosion, we can treat the distribution $f_{LL}$ as the extreme
case (the Maxwellian form then proves in between this distribution
and the delta-function-like form).

In the case of the collapse into a BH (i.e. when the
presupernova mass $M_*>M_{cr}=35 M_\odot$), we assume that
the kick velocity is proportional to the mass lost during the collapse:
\begin{equation}
w_{bh}=w\frac{1-k_{bh}}{1-\frac{M_{OV}}{M_*}}
\end{equation}
where $M_{OV}=2.5$M$_\odot$ is the Oppenheimer-Volkoff limit for NS
mass. This function satisfies boundary conditions
$w_{bh}=0$ at $k_{bh}=1$ (when the total mass of the collapsing
star goes into a BH) and $w_{bh}=w_{ns}$ once $M_{bh}=M_{OV}$.

\section{Effect of the kick velocity on the
binary pulsar populations}

Let us consider what fraction of different types of binary radiopulsars
can be obtained within the framework of the modern evolutionary
scenario for binary stars with the supernova
collapse anisotropy included.  Several attempts of such kind have
been made over the last ten years (Kornilov \& Lipunov 1984; Tutukov
et al. 1984; Dewey \& Cordes 1987; Bailes 1989).  The existing at that
time observational data convicingly pointed to a presence of a small
kick velocity of about 70-100 km/s.  However, the statistics of binary
pulsars at that time was very poor.  Of course, all such studies
are restricted to considering the evolution of an ensemble of stars
initially originated as binaries and not formed due to tidal capture
in globular clusters.

We wish to compare the calculated and observed numbers of binary
radiopulsars with neutron stars (PSR+NS), white dwarfs (PSR+WD),
planets (PSR+PL), and normal OB-stars (PSR+OB). Table 1 lists
the observed numbers of
such binary pulsars in the galactic disk and depicts
their assumed origin.

\begin{figure}
\epsfxsize=0.8\hsize
\epsfbox{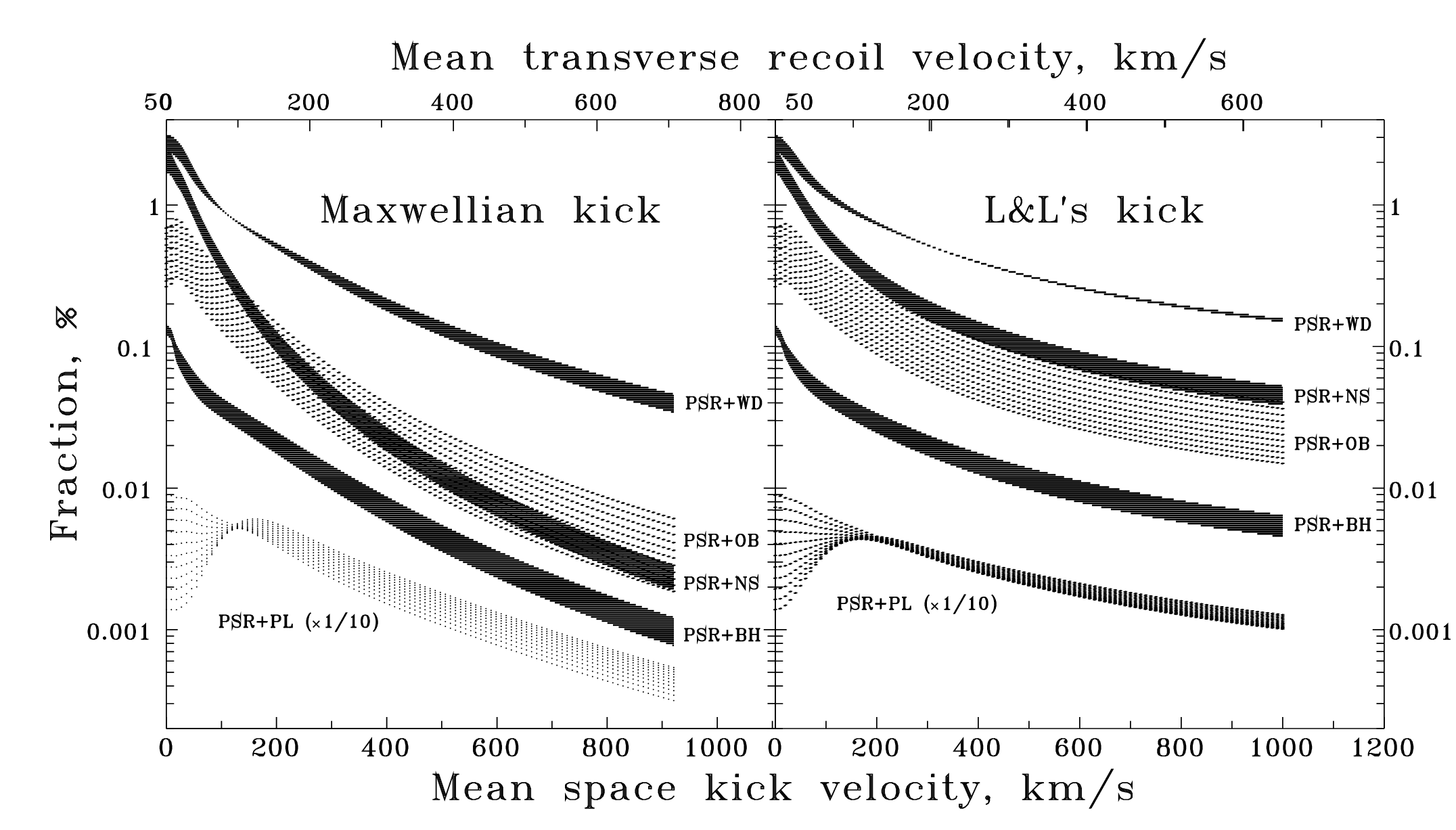}
\caption{
Binary pulsar fractions among the total number of pulsars as
a function
of the mean kick velocity for a maxwellian distribution
(right panel) and Lyne-Lorimer distribution (left panel). The
width of each curve reflects the variation in the power index
$0\le\alpha_q\le 2$ of the initial mass ratio spectrum. The curve for
pulsars+planets is shifted down by one order of magnitude for clarity.
}
\label{zoo1}
\end{figure}

The results of simulation of $10^8$ binaries are presented in
Fig. \ref{zoo1}.  We assumed that in the artificial galaxy we modeled
the number of single stars is equal to that entering binaries, with the
total stellar mass of the galaxy being $10^{11} M_\odot$. The calculations were
performed for the initial mass ratio power $\alpha_q$ ranged from 0 to 2,
which resulted in the notable dispersion of the curves for each
type of binary pulsars seen in Fig. \ref{zoo1}.

Fig. \ref{zoo1} (left panel) shows the dependence of the fraction of
different types of binary pulsars among the total number of
radiopulsars (coming from both single and disrupted binary stars) on
the mean maxwellian kick velocity $w_m$ (three-dimensional).  For
completeness, we added the computed number of PSR+BH binaries (assuming
the critical mass of the pre-collapsing star to be 35 $M_\odot$ and the
mass fraction forming the black hole to be 0.3 after Lipunov et al.
(1994)).  The upper axis shows the mean transverse \it recoil \rm
velocity of pulsars corresponding to the assumed $w_m$. The right panel
shows the same curves obtained for the kick velocity distribution
(\ref{LLkick}) that fits the pulsar transverse velocity distribution
given by Lyne \& Lorimer (1994).  As expected, the increase in the mean kick
velocity decreases all binary pulsar fractions.  In
Fig. \ref{zoo1} the fraction of pulsars with planets (PSR+PL) is
reduced by a factor of 10 for clarity.

The non-monotonic character of the PSR+PL curve is explained by
the fact that generally a small kick velocity causes the new
periastron of the binary system decrease, which makes
dissipative processes more efficient and therefore leads to the white
dwarf Roche lobe overflow with the subsequent planet formation in a wide
binary around a pulsar. At higher kick velocities
the number of PSR+PL gradually decreases.
A more rapid fall of the PSR+NS and PSR+BH curves is caused
by these systems undergoing two collapses (and hence, two kicks).

Obviously, the ``maxwellian'' curves decreases faster (left panel of
Fig. \ref{zoo1}) than the ``L\&L'' ones. This is due to a flat power-law
asymptotics of Lyne \& Lorimer's 2D distribution at low velocities.
Indeed, for large mean kick velocities the maxwellian distribution has
an asymptotic behaviour $f_m\sim w^2$ at slow velocities
($\le 300$ km/s) which are less or comparable with the
characteristic binary orbital velocity, whereas $f_{LL}\sim x^{0.19}$
goes much flatter. We also note that calculations for
delta-function-like kick velocity distribution (not shown in these
figures) naturally lead to even faster drop of the binary
pulsar fractions.

\begin{figure}
\epsfxsize=0.8\hsize
\epsfbox{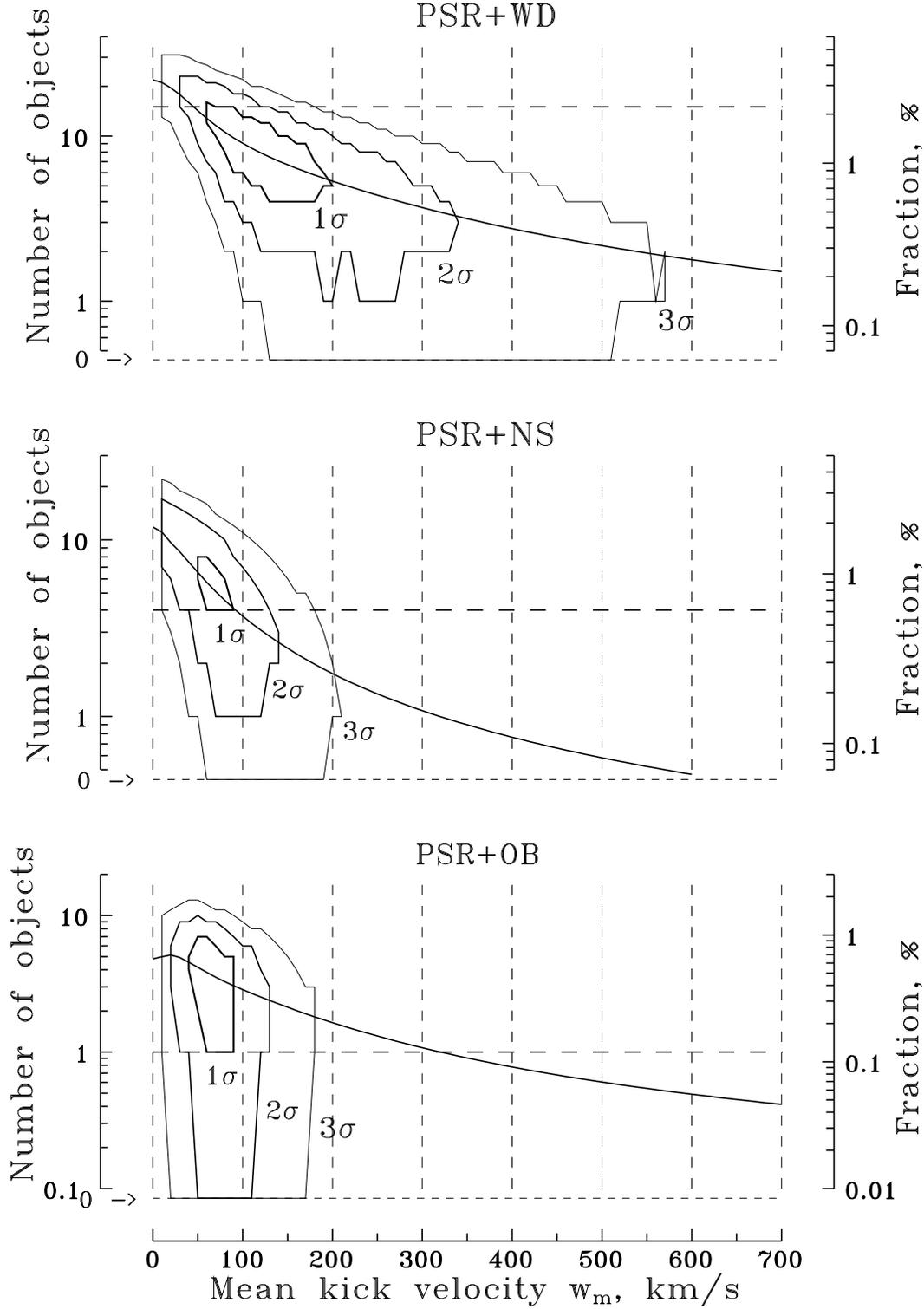}
\caption{
Number and fraction of PSR+WD (a), PSR+NS (b) and PSR+OB (c)
systems (for $\alpha_q=0$)
as a function of the mean kick velocity distributed according
to the law (\protect\ref{LLkick}) that produces pulsars
transverse velocities fitting
the Lyne \& Lorimer's data. Shown also are $1\sigma$, $2\sigma$ and
$3\sigma$ errors computed assuming Poissonian statistics of binary
pulsar numbers among the total number of visible pulsars. The observed
fractions are shown by the dashed line.
}
\label{zoo2}
\end{figure}

Fig. \ref{zoo2} shows the computed fractions of PSR+NS, PSR+WD and PSR+OB
among the observable pulsars in the particular case of
$\alpha_q=0$ (flat intitial mass ratio distribution) assuming kick
velocity distribution $f_{LL}$ given by Eq.(\ref{LLkick}). As the
fractions are small, we can apply Poissonian statistics to evaluate the
significance of the calculations. We draw the confidence levels
corresponding to 68.4\%  (1$\sigma$), 95.45\%  (2$\sigma$) and 99.73\%
(3$\sigma$) probabilities. As is seen from the figure, in all three
cases the observed numbers (dashed lines) fall within 3$\sigma$ errors
only for the mean kick velocities less than $\approx 200$ km/s.

\begin{figure}
\epsfxsize=0.8\hsize
\epsfbox{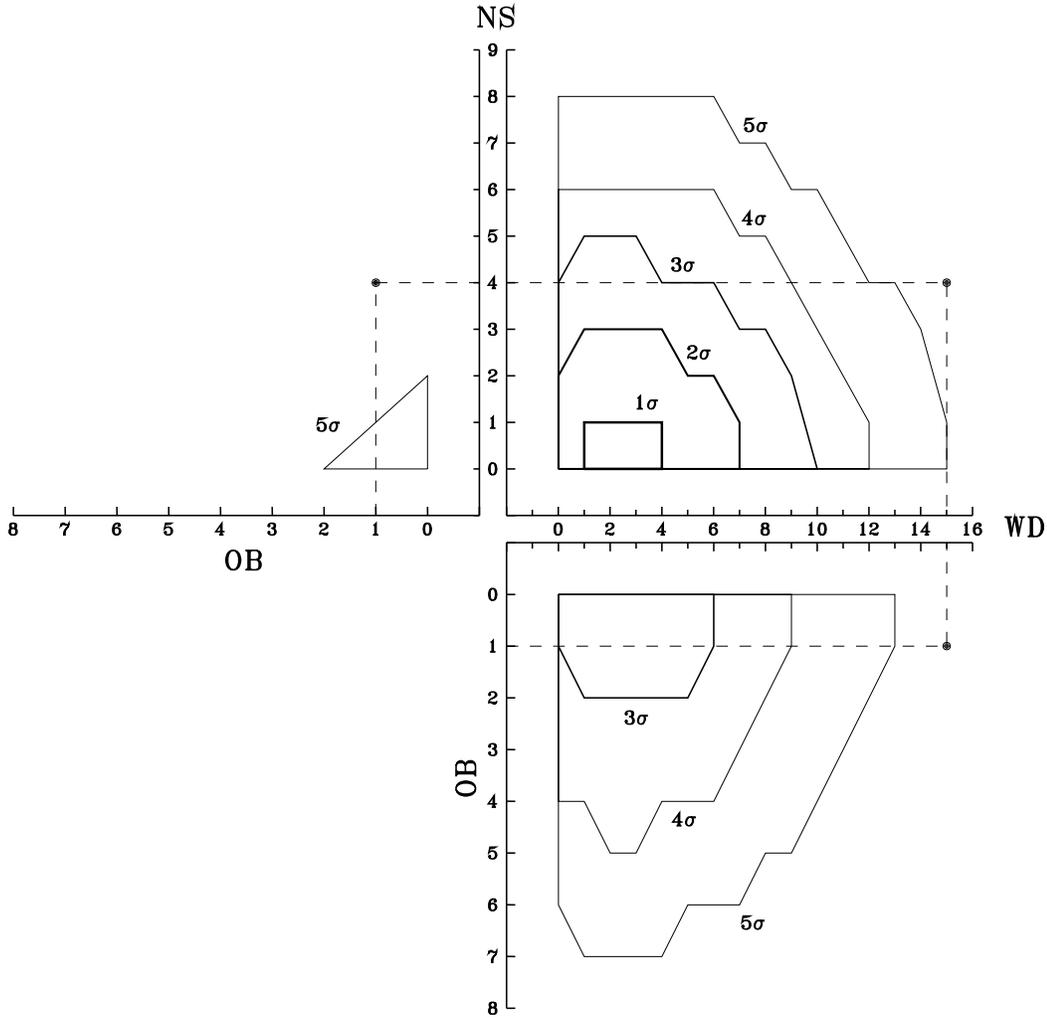}
\caption{
The cross-sections of `phase space'  by the plane
coming through the calculated point corresponding to Lyne \& Lorimer's
3D-kick (\protect\ref{LLkick}) at $w_0=400$ km/s. Solid lines show boundaries of the phase
volume corresponding to 1-5$\sigma$ confidence level.  The numbers of
objects are normalized so as to have 600 single visible pulsars in the
modelled galaxy. The observed numbers from Table 1 are shown by the
dashed lines.
}
\label{zoo3}
\end{figure}

In Fig. \ref{zoo3} we present the `phase space' cuts
coming through the calculated point for
Lyne-Lorimer 3D-kick (\protect\ref{LLkick}) with $w_0=400$ km/s.
The solid lines
show the phase volume boundaries corresponding to 1-5$\sigma$
confidence levels.  The numbers of objects are normalized so as to have
700 single visible pulsars in the modeled galaxy. The observed numbers
from Table 1 are shown by the dashed lines.  The figure demonstrates
that the mean kick velocity 400 km/s lies far outside 5$\sigma$
level for all types of objects.
We find that the combined calculated and observed
numbers differ by less than 3 times for the mean kick velocities
$w_m<200$~km/s at $\approx 100$ km/s.

Let us turn now to the observable 2D distribution of the pulsar
transverse velocities (Lyne \& Lorimer 1994; Frail et al. 1994). As
mentioned above, it is well reproduced by the law
(\protect\ref{LLkick}) with $w_0=400$ km/s, which corresponds to the
mean velocity $w_m\simeq 0.83 \times w_0 \approx 332$ km/s (we note
that the mean recoil space (3D) velocity of pulsars with such kick
distribution is $\approx 370$ km/s; obviously, for high velocities kick
and recoil velocities coincide with each other). The mean kick velocity
that high would correspond to a five-fold disagreement with the
observed situation.  Higher mean recoil transverse velocities reported
by Frail et al.  (1994), 500-1000 km/s, would lead to a more dramatic
inconsistency.  Fig. \ref{zoo1} clearly demonstrates that for high mean
space velocities $> 300$ km/s the observed binary pulsar ``Zoo'' would
rapidly vanich.

As the computed binary pulsar fractions are strongly dependent on the
kick velocity, a question arises as to how accurately we derive them
from observations.  The fraction of binary pulsars among the total
single pulsar population seems to be minimally subjected to possible
selection effects. Indeed, let us compare the observable fraction of
binary pulsars (Table 1) with pulsar birthrates and galactic numbers
derived from pulsar statistics with account of a number of selection
effects (Lorimer, 1995; Curran \& Lorimer, 1995).  For example, Curran
\& Lorimer (1995) estimate the total number of the potentially
observable (i.e. with no account of beaming) NS+PSR pairs in the Galaxy
$\sim 240$.  The total number of the potentially observable galactic
single pulsars (Lorimer et al., 1993) is estimated to be $1.3\pm
0.2\times 10^4$. Hence, the galactic fraction of PSR+NS to PSR is
$240/13000\approx 1.8\%$ with a half-order accuracy, and the observed
ratio of $0.5\%$ (Table 1) thus places an \it upper limit \rm
to the kick velocity.
We note that PSR+NS systems are most sensitive to the kick velocity
(because for them the kick effect doubles) and, therefore, they are the
most important for our consideration.

The number of PSR+WD systems is less accurately known. For example, if
we use the ratio of local birthrate of low-mass binary pulsars
$2-4\times 10^{-9}$ kpc$^{-2}$ yr$^{-1}$ (Lorimer, 1995) to the local
pulsar burthrate $6-12 \times 10^{-6}$ kpc$^{-2}$ yr$^{-1}$ (Lorimer et
al. 1993), we would obtain the fraction PSR+WD/PSR $0.4-6\%$, assuming the
average age of PSR+WD pulsars 100 times that of single PSR.  While this
figures agrees with our estimate $\sim 2.1\%$ (Table 1), they are very
uncertain.  Nevertheless, Fig. \ref{zoo1} demonstrates that the
observed NS+WD statistics does not contradict an optimal kick velocity
value of 150-200 km/s.

\begin{figure}
\epsfxsize=0.8\hsize
\centerline{\epsfbox{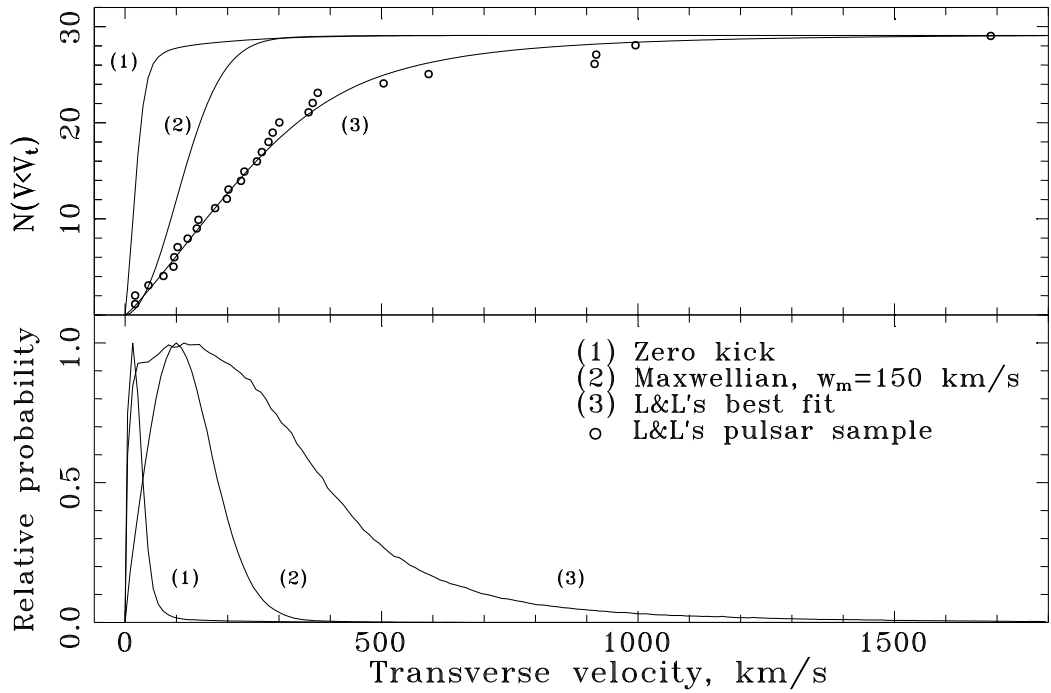}}
\caption{
The simulated pulsar transverse velocity distribution (bottom
panel) and cumulative distribution (upper panel) for different kicks.
The dots are the observed cumulative distribution from Lyne \& Lorimer (1994).
}
\label{zoo5}
\end{figure}

To conclude this Section, in Fig. \ref{zoo5} we present the calculated
pulsar transverse recoil velocity distribution for three specific
cases: with zero kick (curve (1)), the maxwellian kick with $w_m=150$
km/s and distribution (\ref{LLkick}) best-fitting Lyne-Lorimer data
(shown by filled circles).  The cumulative distribution is reproduced
for comparison (upper panel). Clearly, the maxwellian distribution
is unable to reproduce the high-velocity tail observed in the observed
pulsar transverse velocity.

\section{Fraction of active pulsars in relativistic
compact binaries}

Before proceed to studying the effect of kick velocity on the rate of
binary NS/BH mergings, we wish to answer the question:  why estimations
of merging rates obtained from pulsar statistics are systematically by
two orders of magnitude less than those obtained from evolutionary
calculations?  A possible answer proposed by van den Heuvel (1992)
and Tutukov \& Yungelson (1993) is that many relativistic compact
binaries are formed after the common envelope stage with too short
orbital period and hence coalesce on a short time-scale.
In our opinion, this is not the case.
Our calculations (Lipunov et al., 1996c) and integration
of the curve for NS+NS birthrates presented in Fig. 1 of
Tutukov \& Yungelson's paper show that  binary NS+NS with life-time before
coalescence less than $10^7$ years contributes only $\sim 10-20\%$ to
the total galactic merging rate.

We propose a more clear explanation connected
to the physical state of the rotating NS, becacuse not all NS in relativistic
binaries are shining as pulsars. To estimate this effect a thorough
treatment of NS spin evolution is needed. The major uncertainty
here is whether the magnetic field of NS decay or not.

\begin{figure}
\centerline{%
\epsfysize=0.5\hsize%
\epsfbox{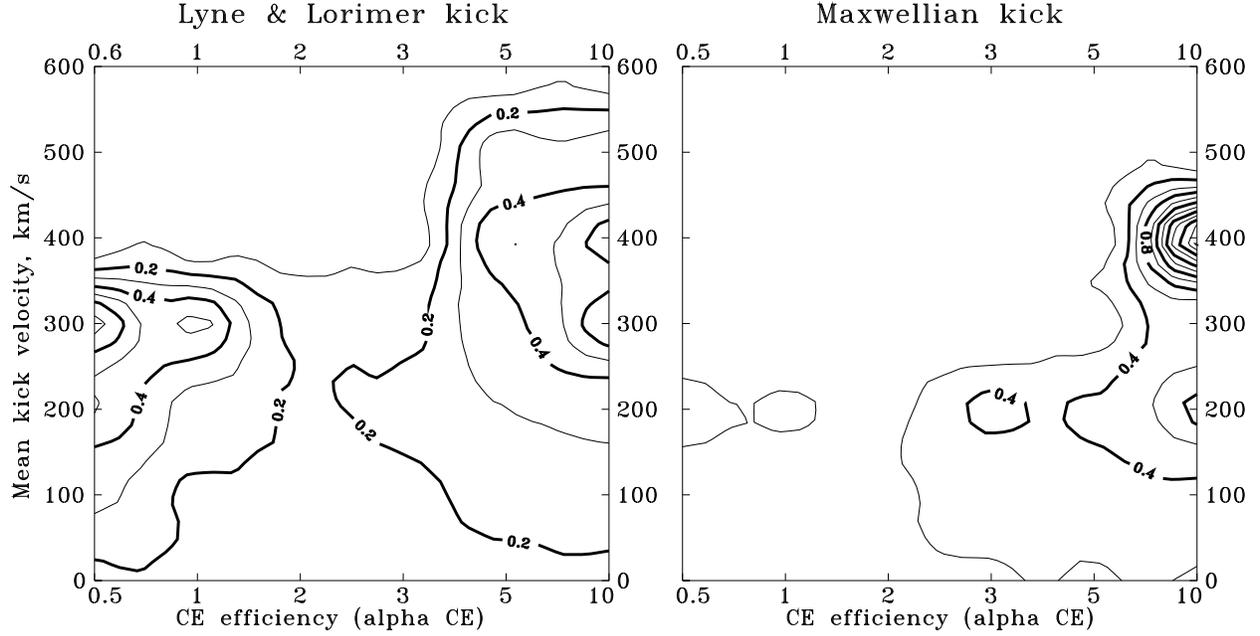}}
\caption{The percentage of binary pulsars with NS companions
among all NS+NS binaries as a function of the common envelope
efficiency $\alpha_{CE}$ and mean kick velocity $w_m$ for
maxwellian (right panel) and Lyne-Lorimer kick velocity
distributions assuming an exponential magnetic field decay
on the characteristic timescale $t_d=10^7$ years}
\label{ns1}
\end{figure}

\begin{figure}
\centerline{%
\epsfysize=0.5\hsize%
\epsfbox{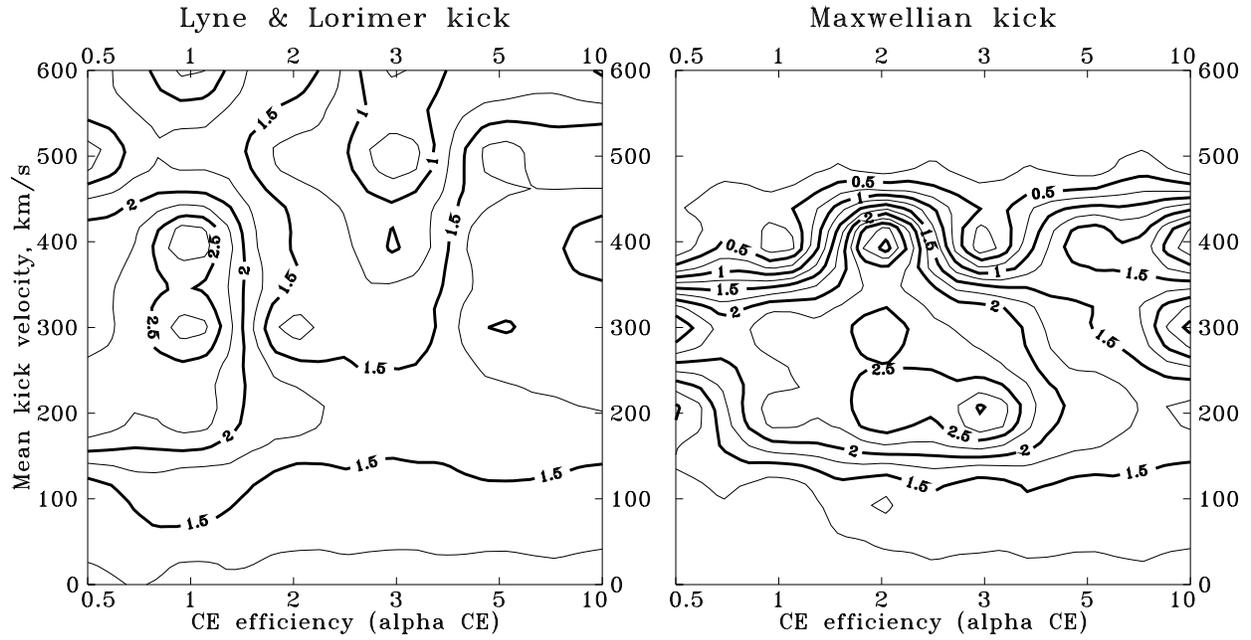}}
\caption{The same as in Fig. \protect\ref{ns1} for $t_d=10^8$ years}
\label{ns2}
\end{figure}

Fig. \ref{ns1}-\ref{ns2} show the fraction of active binary pulsars
with NS component among the total number of binary NS in the whole
range of orbital periods for two values of the exponential magnetic
field decay time $10^7$ and $10^8$ years.  The "low mass-loss" scenario
for stelar wind was used.  The calculated fraction is only slightly
dependent on the binary orbital period and thus can be considered as an
underestimation factor of binary NS merging rate.  As seen from
Fig. \ref{ns1}-\ref{ns2}, this factor is typically of order 1 -- 3\%
for a wide range of the mean kick velocities and common envelope
efficiencies. Combined with beaming factor, this brings the
underestimation factor to less than 1\%.  Thus, the lower limits to the
binary NS coalescence rate derived from binary pulsar statistics must
be increased by a factor of 100, which reconciles them
with theoretical expectations.

\section{Effect of the kick velocity on relativistic binary
merging rates}

In view of the importance of the kick velocity for
evolutionary scenario, in this section we focus on
how kick velocity affects the double NS/BH merging rates.

\subsection{Low stellar wind mass loss}

\begin{figure}
\centerline{%
\epsfysize=0.9\vsize%
\epsfbox{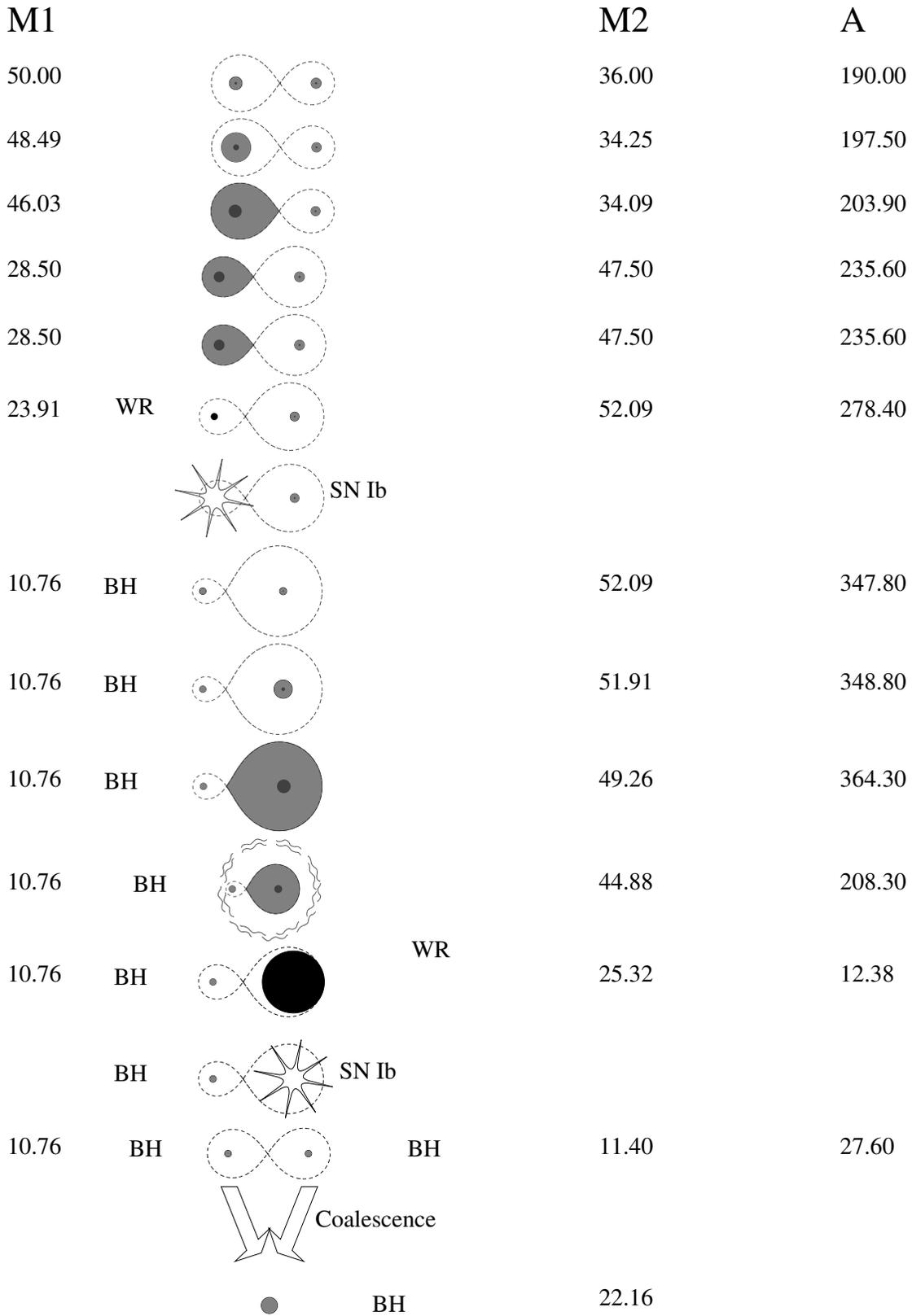}}
\caption{Example of a binary BH system formation}
\label{ns_2B}
\end{figure}

First we consider the low stellar wind mass-loss evolution.  We start with
reproducing a typical evolutionary track leading to the formation of
a coalescing binary BH (Fig. \ref{ns_2B}).
BH formation parameters are $M_*=35
M_\odot$ and $k_{bh}=0.3$. Zero kick velocity is used to compute this
track.
Each evolutionary stage shown in these
figures is marked with masses of components and orbital separation (in
solar units).
(Everyone can try to construct these and many other
tracks under different assumption via WWW-server at the URL {\tt
http://xray.sai.msu.su/sciwork/scenario.html}).

\begin{figure}
\centerline{%
\epsfysize=0.5\hsize%
\epsfbox{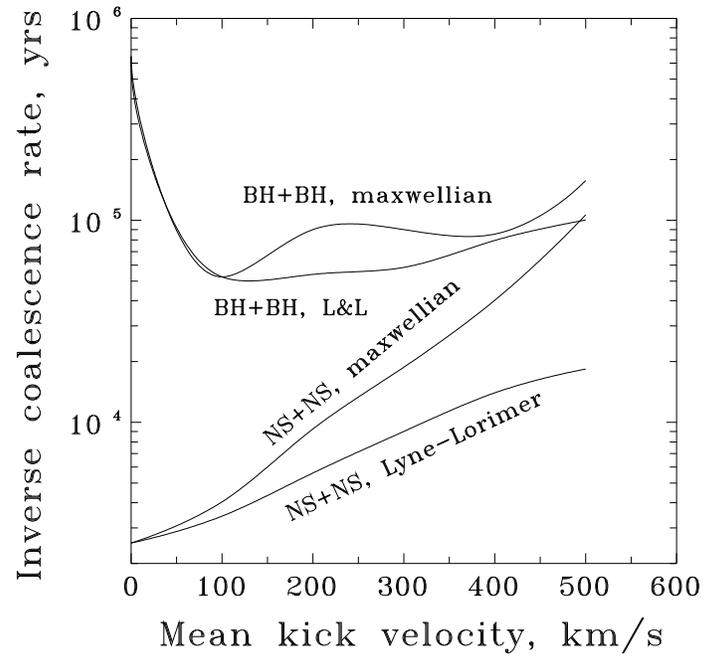}}
\caption{The total merging rate of NS+NS and BH+BH binaries
for different kick velocity laws
for $t_d=10^8$ yr and assuming low stellar wind mass-loss.}
\label{ns3}
\end{figure}

Fig. \ref{ns3} shows merging rates of the relativistic compact binaries
as a function of the mean kick velocity assuming different kick
distributions (maxwellian or Lyne-Lorimer).  The stronger
decrease in the case of the maxwellian distribution is due to the
steeper dependence ($\sim x^2$) of this distribution at lower
velocities.  Notice also an increase in the NS+BH merging rate at small
velocities and its constancy at higher velocities. From Fig.
\ref{ns3} we see that the theoretical expectation for the NS+NS merging rate
in a model spiral galaxy with the total stellar mass of $10^{11}$ M$_\odot$ lie
within the range from $\sim 3\times 10^{-4}$ yr$^{-1}$ to $\sim
10^{-5}$ yr$^{-1}$, depending on the assumed mean kick velocity and the
shape of its distribution. For zero kick velocity,
our results are fully consistent with earlier estimates
(Lipunov et al. 1987, Tutukov \& Yungelson, 1993).

For  Lyne-Lorimer kick velocity law with the mean
value of 400 km/s, we obtain $R_{NS+NS}\approx 5\times 10^{-5}$
yr$^{-1}$, which again yeilds the estimate of 1 NS+NS merging per $\sim
10,000$ years considering the beaming factor.  Note that the event rate
of $10^{-5}$ mergings per year at $w=450$ km/s was recently obtained by
Portegies Zwart \& Spreeuw (1996) who assumed a maxwellian distribution
for kick velocity. This figure is in line with our calculations (upper
line in Fig. \ref{ns3}). We repeat, however, that the maxwellian
kick velocity distribution would be in a strong disagreement with
binary pulsar fractions even at low kick velocities (see Fig. 1).

Our galactic NS+NS merging rate ($5\times 10^{-5}$ yr$^{-1}$) is also
exceeds the so-called "Bailes upper limit" to the NS+NS birthrate in
the Galaxy $\sim 10^{-5}$ yr$^{-1}$ (Bailes, 1996).  Indeed, following
Bailes, the fraction of "normal" pulsars in NS+NS binaries which
potentially could merge during the Hubble time among the total number
of known pulsars is less than $\sim 1/700$. Multiplying this by the
present birthrate of "normal" pulsars $1/125$ yr${-1}$ (Lorimer et al. 1993),
we obtain this limit $10^{-5}$ yr$^{-1}$. In contrast, our NS+NS
merging rate is 5 times higher for Lyne-Lorimer kick velocity
distribution even at $w=400$ km/s.  First we note that the accuracy of
the Bailes limit is a half-order at best; in addition, the pulsar
birthrate 1/125 yr$^{-1}$ is a lower limit and
at least 4 times lower than the birthrate
of massive stars ($>10 M_\odot$) which produce neutron stars in our
Galaxy (once per 30 years according to Salpeter mass function).

This discrepancy may be decreased considering the existing uncertainty
in the pulsar beaming factor and, which may be more important, the
still-present uncertainty in pulsar distance scale which influences the
estimate of the total galactic number of pulsars and hence their
birthrate.

In addition, one more important point indicates that the Bailes limit
cannot be universal. As our calculations of binary NS+NS merging rate
evolution show (Fig. 1 in Lipunov et al., 1995), binary NS formed
as long as 10-15 billion years ago (`relic' NS binaries) may coalesce
at the present time and their merging rate may exceed the Bailes limit, so
the true merging rate should be determined by baryon fraction
transformed into the first population stars and has no connection with
the currently observed pulsar statistics (the Bailes limit).

Two details from Fig. \ref{ns3} worth noting: 1) the binding effect at small
kick velocities and 2) the smaller effect of high kicks on the BH+NS (BH)
rate.  The first fact is qualitatively clear: a high kick leads to the
system disruption; however, if the system is survived the explosion,
its orbit would have a periastron distance
always smaller than in the case without
kick. During the subsequent tidal circularization a closer
binary sytem will form which will spend less time before the merging.
The binding effect of small recoil velocities is very pronounced in the
case of binary BH. At higher kicks their merging rate decreases slower
due to higher masses of the components.

\subsection{High stellar wind mass loss}

Now we turn to the high stellar wind mass-loss evolution.
We recall that within the framework of this scenario,
single massive stars lose most their mass by stellar wind
to leave a WR star of 8-10 M$_\odot$ and even smaller
which then explodes presumably as a SN type Ib and leaves NS or BH as
a remnant. Here only one BH formation parameter is needed --
$k_{bh}$.

\begin{figure}
\centerline{%
\epsfysize=0.5\hsize%
\epsfbox{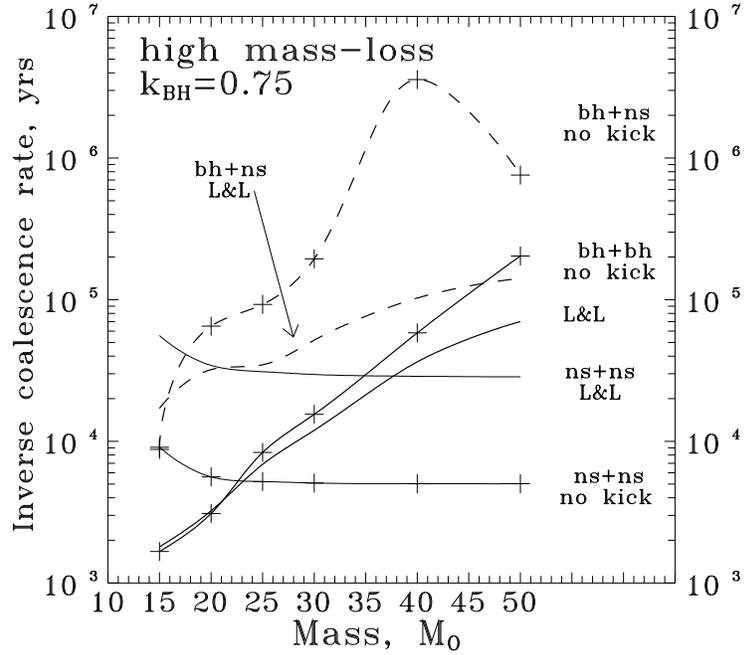}}
\caption{The
merging rate of NS+NS, NS+BH and BH+BH binaries for different kick
velocity assumptions for the high stellar wind mass-loss scenario as a
function of the minimum mass of  main-sequence star producing BH in the
end of evolution.}
\label{ns_high}
\end{figure}

The results of computing galactic merging rates of binary relativistic
stars for $k_{bh}=0.75$ and different kicks (no kick or Lyne-Lorimer kick
with the mean velocity $w=400$ km s$^{-1}$ are presented in Fig.
\ref{ns_high}. The rates are plotted against the minimum initial mass
of main-sequence star that evolves to form a BH.

\section{Relative ratio of binary
BH and NS merging rates}

It is widely recognized that binary BH mergings may be
as important as binary NS mergings.
Let us make a crude estimation
of the BH binary merging rate.  Within the framework of our assumptions,
the mass of BH progenitor is $35 M_\odot$ (if low mass-loss
stelar wind is effective), which corresponds roughly to
$M_{ms}\sim 60 M_\odot$ on the main sequence (we recall that according
to  the evolutionary scenario, mass of a star after mass transfer is
$M_{core}\simeq 0.1 M_{ms}^{1.4}$).  On the other hand, any star with
$M_{ms}\ge 10 M_\odot$ evolves to form a NS.  Using the Salpeter mass
function ($f(M)\propto M^{-2.35}$), we obtain that BH formation rate
relates to NS formation rate as $(60/10)^{-1.35}\approx 0.09$.
Extrapolating this logic to binary BH/NS systems, we may expect
$R_{bh}/R_{ns}\sim 1/10$, within a half-order accuracy. Actually, the
situation is complicated by several factors:
the presence of the kick velocity during
supernova explosion which may act more efficiently in the case of NS
formation; mass exchange between the components; distribution
by mass ratio, etc. Our calculations account for
most these factors.

\begin{figure}
\centerline{%
\epsfysize=0.5\hsize%
\epsfbox{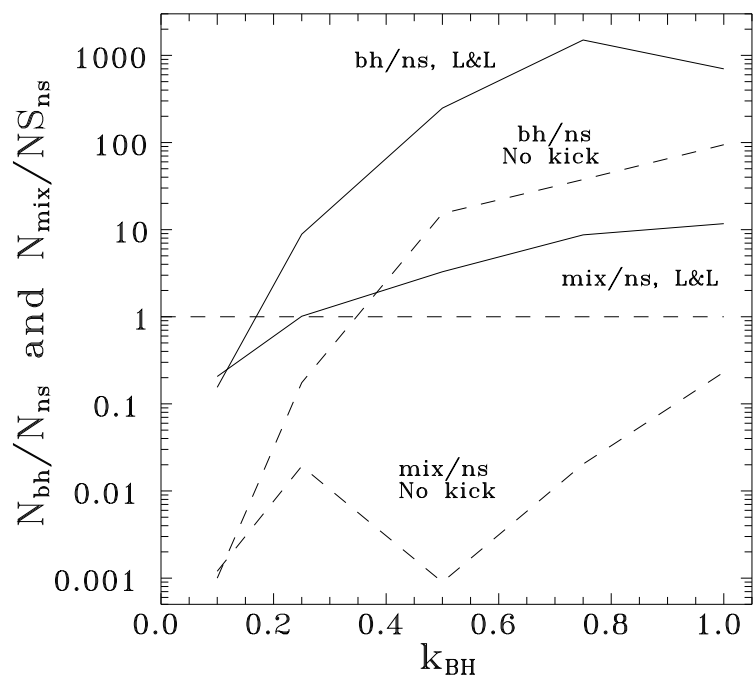}}
\caption{The ratio of NS+BH and BH+BH mergings relative to NS+NS events
detected by a gravitational wave detector with a given sensitivity, as
a function of parameter $k_{bh}$
assuming low stellar wind mass-loss scenario.
Calculations with Lyne-Lorimer kick velocity distribution
and without kick are shown by solid and dashed lines,
respectively.}
\label{ns4}
\end{figure}

\begin{figure}
\centerline{%
\epsfysize=0.5\hsize%
\epsfbox{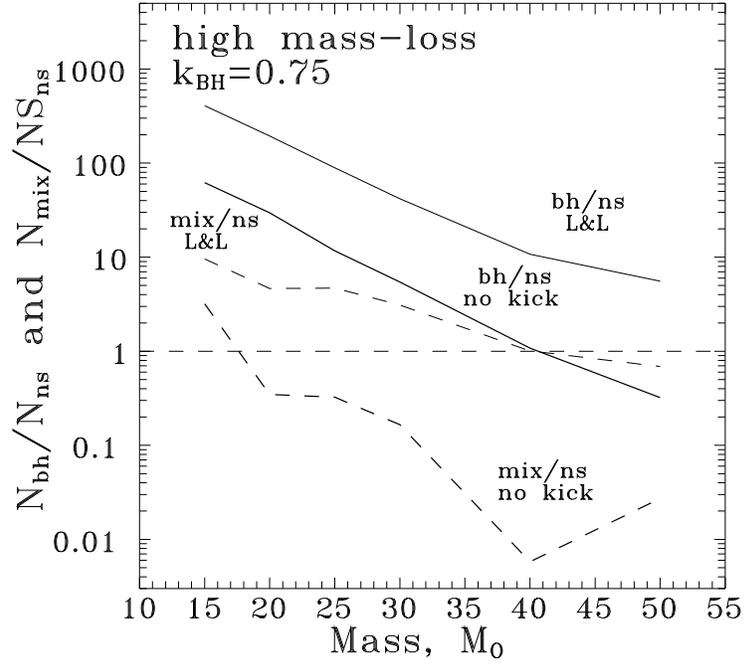}}
\caption{The ratio of NS+BH and BH+BH mergings relative to NS+NS events
detected by a gravitational wave detector with a given sensitivity, as
a function of minimal mass of a main sequence star producing black hole
for different kick velocity laws (not specified for NS+BH and BH+BH
binaries in the figure) within the framework of high stellar
wind mass-loss scenario.}
\label{ns5}
\end{figure}

Fig. \ref{ns4}-\ref{ns5} demonstrate the relative detection rate of
coalescing BH and NS  binaries obtained by a gravitational
wave detector with a given sensitivity. In Fig. \ref{ns4} we present
the relative detection rate assuming the low stellar wind mass loss scenario
as a function of BH-formation parameter $k_{bh}$,
the fraction of the stellar mass that forms a BH
after the collapse, for two particular initial distirbution of
binaries by mass ratio: $\alpha_q=2$, a strongly peaked
toward equal initial masses distribution (as follows from the
independent study of X-ray binaries modeling, LPP 1996a), and
$\alpha_q=0$, a flat initial mass ratio distribution. These
calculations were
performed assuming both zero kick velocity during supernova
explosion (the dashed line in the figure) and Lyne \& Lorimer's kick
velocity distribution with the mean velocity $w_0=400$ km/s.
It is seen that the form of the initial mass ratio distribution only
slightly affects the results.  Thus  the ratio of detection rates for
merging BH/NS binaries (\ref{rates}) may well exceed unity for a wide
range of parameters.

In Fig. \ref{ns5} we plot the same ratio of BH to NS
merging events as in Fig. \ref{ns4}, but calculated  for
the high stellar wind mass-loss scenario. The parameters
are the same as for caculations shown in Fig. \ref{ns_high}
($k_{bh}=0.75$, Lyne-Lorimer kick with $w=400$ km s$^{-1}$).
This ratio is plotted  as a function of the minimal initial mass of
main-sequence star capable of producing a BH in the end
of its evolution.

\section{Duscussion}
\subsection{Uncertainties of the results}

Three independent groups of factors can potentially affect the results
of evolutionary simulations:
(1) the assumptions underlying the scenario of
binary evolution used; (2) the intrinsic accuracy of Monte-Carlo
simulations and (3) selection effects.

(1) Scenario assumptions. The present scenario for the binary star evolution
has many parameters, some of which has a purely theoretical sense. Of them,
the most crucial are the shape of the initial mass ratio distribution
and the kick velocity imparted to neutron stars at birth.
While the former was explored in Lipunov et al. (1995b), the
latter is studied in the present paper for a wide range of possible
kick velocity distributions. From the point
of view of the effect the kick has on the binary system evolution,
the most crucial proves the low velocity
tail of the distribution. Having chosen the Maxwellian (\ref{Mxw})
and Lyne \& Lorimer laws (\ref{LLkick}), we thus studied power indexes
of the low velocity asymptotics from the range 0.19-2.

Another important question is how the magnetic field of neutron stars
is distributed and how it evolves.  Unfortunately, the initial magnetic
field distribution for neutron stars cannot be directly taken from
pulsar observations because of strong selection effects for highest and
lowest magnetic fields, so in the present calculations we have chosen
the broadest possible initial magnetic field distribution
($d\log\mu=const$). Then, we assumed exponential field decay.  The
results shown in Fig. \ref{ns1}-\ref{ns4} were obtained for a decay time of 100
Myr. Taking $\tau=10^7$ yr does not alter appreciably the calculated
fractions of binary pulsars (especially PSR+NS), unless one assumes a
strong dependence of pulsar magnetic fields on the evolutionary history
(see Camilo et al. 1994). This may have effect on the pulsars
originated from low-mass binary systems.  However, the example of
Her X-1, which provides evidence of a high magnetic field in an
accreting low-mass binary system, shows that one should consider such
possibilities with caution, and in the present work this complication
of the evolutionary scenario is not taken into account.

At last, for very massive stars from which BH are formed
the crucial is how fast stellar wind mass loss occurs.
We specially
used both high and low stellar wind mass-loss cases when
studying BH formation as two extremal scenarios for binary
evolution.

(2) The intrinsic accuracy of Monte-Carlo simulations is determined by
the number of trials which in our case is $10^6$ per each run, so expected
errors are much less than the scenario uncertainties.

(3) Selection effects. These are important for comparison with
observations.  In estimating the ratios of binary pulsars, one may
argue that different selections affect the numbers of single and
binary pulsars (e.g. Johnston \& Kulkarni 1990), for example, that the
recycled pulsars may constitute a larger fraction among single pulsars
than is actually observed. In this case, however, we use a \it lower
limit \rm of the actual binary pulsar fractions, and thus obtain \it
conservative \rm results, as increasing the observed fractions of
binary pulsars would aggravate the problem of high mean kick velocities
implied by recent pulsar observations (Fig. \ref{zoo1}).
When we derive binary
pulsar mass fractions using pulsar birthrates given
by Lorimer et al. (1993), Lorimer (1995) and Curran \& Lorimer (1995)
(especially for NS+PSR fraction),
we again obtain even smaller kick velocity $\sim 80-100$ km/s.

\subsection{Does the nature require an ad hoc asymmetric kick?}

In the recent paper, Iben \& Tutukov (1996) try to explain
self-consistently the observational pulsar data without
invoking an ad hoc kick velocity imparted to neutron star at birth.
They criticize the Lyne \& Lorimer's data
referring to the pulsar distance scale uncertainties. In fact,
the Lyne \& Lorimer's distribution has two characteristic features:
(1) a very high \it mean \rm velocity (at least two times higher than
that was believed earlier) and (2) an unusually long power-law tail
of pulsar velocities (in contrast to, for example, the exponential
tail in a Maxwellian distribution). Our consideration shows that
the high mean space velocity of pulsars is in a paradoxical contradiction
with the observed Zoo of binary pulsars. In this sense, our conclusions
coincide with Iben \& Tutukov's ones (no high kick velocity). However,
we insist on that if the mean kick velocity decreases keeping the
qualitative behavior of the pulsar velocity distribution (i.e.  the
power-law high-velocity tail, which does not relate to the assumed
distance scale and therefore does not depend on selective effects of
this kind), we have to invoke the kick hypothesis.

This fact, as well as the best coincidence between observations
and calculations
at $w\sim 100-200$ km s${-1}$, indicate the need for a small natal
kick velocity, as was previously shown by Kornilov \& Lipunov (1984).
We recall that in the latter paper the authors relied upon the lack of
the observed binary pulsars with OB-stars at that time, and this
difficulty could be overriden by assuming, as Iben \& Tutukov (1996)
do, that the pulsar phenomenon appears only for neutron stars born in
close binary systems with orbital periods smaller than $\sim 10$ days.
But in the present paper we use all variety of the observed
pulsars and, what is especially important, their relative number
with respect to single pulsars, which in no ways can be reduced
within the framework of the hypothesis that
no pulsars appears in wide binary systems. Therefore, our statistical
analysis of pulsar numbers, together with the observation of the
high-velocity power-law tail in pulsar velocity distribution,
makes the assumption of the presence of a moderate kick velocity
quite realistic.

In addition to the statistical arguments, there is a lot of independent
observational indications of the moderate kick velocity existence.  For
example, the observed inclination of the gaseous disk around Be-star in
PSR B1259-63 relative to the orbital plane (Melatos et al. 1995) could
hardly arise without supernova explosion asymmetry.  Many observed
long-term variability in classical X-ray binaries (such as Her X-1, SS
433, LMC X-4, etc., see Cherepashchuk (1981)) may be quantitatively
expalined in a natural way if the first supernova explosion in these
systems was asymmetric.  At last, a direct evidence for a kick velocity
of at least 100 km/s has recently been obtained from observations of
precessing binary pulsar orbit in PSR J0045-7319 in the SMC (Kaspi et
al. 1996).

\section{Conclusion}

The purposes of the paper was to study the effect of kick
velocity (1) on the observed numbers of binary pulsars with different
companions and (2) on the the expected merging rates of binary
relativistic stars -- double NS, NS+BH and double BH systems.
\vskip\baselineskip
1. The calculations presented above demonstrate a strong decrease in
number of binary pulsars with diverse secondary components as a
function of the assumed kick velocity imparted to a neutron star at
birth. On the other hand, high birth velocities of pulsars seem to be
really observed (Lyne \& Lorimer 1994; Frail et al. 1994).  Thus, a
paradoxical situation emerges: we have to make a choice between the
high birth space velocities for the pulsars and the observed reach Zoo
of binary pulsars in the galactic disk!  Unless the observational data
are subjected to a number of selection effects leading to an
underestimation of low-velocity pulsars (e.g. Tutukov et al. 1984), a
possible outcome from this paradoxical situation could be a slow steady
acceleration of neutron stars which would not disrupt the binary
system, but would impart a high space velocity to single pulsars (for
example, a rocket mechanism by Tademary \& Harrison (1975) caused by an
asymmetric pulsar emission).  However, the evidence for high
velocities of young pulsars associated with supernova remnants (Frail
et al. 1994) and the lack of a firm evidence for the velocity
increasing with age are not consistent with this mechanism.  The
presence of numerous pulsars in globular clusters also put constraints
on the efficiency of such a mechanism.

What to do with the fast pulsars moving with velocities $>1000$ km/s?
The highest pulsar velocities resulted from the disruption of a neutron
star -- massive helium star binary during a spherically-symmetric
collapse of the helium star cannot be greater than $\sim 700-800$ km/s.
Recently suggested mechanism by Bisnovatyi-Kogan (1993)
allows the birth pulsar velocities as high as 3000 km/s, but still requires a
very strong magnetic field ($B\gg B_{cr}\simeq 4.4\times 10^{13}$~G) at
birth; Imshennik's (1992) mechanism of double neutron star formation
during the collapse with the subsequent Roche lobe overflow of the less
massive neutron star and its explosion after a minimum mass of about
0.1 M$_\odot$ has been reached, although does not need the magnetic
field that high, could lead to very high individual space velocities,
but hardly to the high \it mean \rm kicks required to fit the
observations. A promising could also be the mechanism of
anisotropic neutrino emission during the supernova explosion
recently suggested by Burrows \& Hayes (1995). These mechanisms
will be checked after the gravitational waves detection will have been
possible by future LIGO/VIRGO experiments.

We conclude that the observed diversity and the number of the binary
pulsars, while do require a small kick velocity ($\sim 100-200$ km/s)
to be present, are, at the same time, in a strong contradiction with
the presently derived high mean transverse velocities of pulsars.

2. We also studied how the merging rate of relativistic
binary compact stars (NS+NS, NS+BH, BH+BH), which are amongst the
primary targets for future LIGO type gravitational wave detectors,
depends on some crucial parameters of the modern evolutionary scenario
of binary stellar systems, namely on the common envelope eficciency
$\alpha_{CE}$ and mean recoil velocity imparted to NS at
birth. Assuming the Salpeter law for primary mass, a flat initial
binary mass ratio, a flat magnetic field distribution of young NS and its
exponential decay on a timescale of $10^7-10^8$ yrs, we obtained that
the galactic number of binary NS with orbital periods $P<10$ hrs is at
least a factor of 100 higher than that deduced from binary pulsar
statistics (Phinney 1991, Curran \& Lorimer 1995).  We also obtained
that this rate decreases nearly exponentially with mean kick velocity.
The decrease goes faster in the case of maxwellian disribution for kick
velocity. Currently popular Lyne-Lorimer kick velocity law that yeilds the observed
pulsar transverse velocity distribution would  give the binary NS
merging rate of $5\times10^{-5}$ per year even for extremely
high kick velocity $w=400$ km/s. The merging
rate for NS+BH and BH+BH binaries was found to be about $10^{-5}$ per
year per Galaxy and practically insensitive to the kick velocity for
$w>100$ km/s.

We conclude that the relativistic compact binary merging rate, even for
high mean kick velocities of NS up to 400 km/s, leads to at
least 30-50 such events per year from within a distance of 200 Mpc, which
the LIGO detectors will be sensitive to. The high merging rate of relativistic
binaries with BH implies that it is not excluded that a comparable
(and even probably a major) fraction of all events that will be registered by
LIGO detectors may be due to these events. This urges efforts on
numerical modeling BH+BH coalescences to obtain template
wavefroms from such process.

\section*{Acknowledgements}

The work was partially supported by the INTAS grant No 93-3364, grant
of Russian Fund for Basic Research No 95-02-06053a, grant JAP-100 from
the International Science Foundation and Russian Government and by
Center for Cosmoparticle Physics ``COSMION'' (Moscow, Russia).  VML
thanks Copenhagen University Observatory for hospitality and Prof. Kip
Thorne for stimulating discussions. We also thank anonymous referee
(E.P.J. van den Heuvel) for useful notes, and
Ivan Panchenko for help in visualzing evolutionary
tracks.

{}

\bsp
\label{lastpage}

\clearpage

\clearpage

\clearpage

\clearpage

\clearpage

\clearpage

\clearpage

\clearpage

\clearpage

\clearpage

\clearpage

\end{document}